\documentclass[aps,prl,showpacs,twocolumn,superscriptaddress,floatfix]{revtex4}

\usepackage{graphicx}
\usepackage{dcolumn}
\usepackage{epsfig}
\usepackage{bm}
\usepackage{amssymb}
\usepackage{amsmath}
\usepackage{color}
\newcommand{\gammap}{\dot{\gamma}}
\newcommand{\mms}{~mm\,s$^{-1}$}
\newcommand{\degre}{$^\circ$}
\newcommand{\degC}{$^\circ$C}

\begin{document}

\title{Influence of boundary conditions on yielding in a soft glassy material}

\author{Thomas Gibaud}
\affiliation{Laboratoire de Physique, Universit\'e de Lyon -- \'Ecole Normale Sup\'erieure de Lyon  -- CNRS UMR 5672\\46 all\'ee d'Italie, 69364 Lyon cedex 07, France}
\author{Catherine Barentin}
\affiliation{Laboratoire de Physique de la Mati\`{e}re Condens\'{e}e et Nanostructures, Universit\'e de Lyon -- CNRS UMR 5586\\43 Boulevard du 11 Novembre 1918, 69622 Villeurbanne cedex, France}
\author{S\'{e}bastien Manneville}
\email{sebastien.manneville@ens-lyon.fr}
\affiliation{Laboratoire de Physique, Universit\'e de Lyon -- \'Ecole Normale Sup\'erieure de Lyon  -- CNRS UMR 5672\\46 all\'ee d'Italie, 69364 Lyon cedex 07, France}
\date{\today}

\begin{abstract}The yielding behavior of a sheared Laponite suspension is investigated within a 1~mm gap under two different boundary conditions. No-slip conditions, ensured by using rough walls, lead to shear localization as already reported in various soft glassy materials. When apparent wall slip is allowed using a smooth geometry, the sample breaks up into macroscopic solid pieces that get slowly eroded by the surrounding fluidized material up to the point where the whole sample is fluid. Such a drastic effect of boundary conditions on yielding suggests the existence of some macroscopic characteristic length that could be connected to cooperativity effects in jammed materials under shear.
\end{abstract}
\maketitle

``Soft glassy materials'' constitute an extremely wide category ranging from food products or cosmetics, such as emulsions or granular pastes, to foams, physical gels, and colloidal glasses. One of the most striking characteristics of these materials is the existence of a transition from solidlike behavior at rest to fluidlike behavior when a strong enough external shear stress is applied \cite{Liu:1998,Trappe:2001}. Although defining and measuring the {\it yield stress}, i.e., the stress above which the material becomes fluid, still raise many difficulties, most of them due to thixotropic features resulting from the competition between aging and shear rejuvenation \cite{Barnes:1999}, it is now well established that the yielding transition leads in most cases to {\it shear localization}, i.e., to flow profiles characterized by the coexistence of a solidlike region that does not flow and a fluidlike region that bears some finite shear rate \cite{Pignon:1996,Coussot:2002,Rogers:2008} even in the absence of any geometry-induced stress heterogeneity \cite{Moller:2008}. Such a picture, revealed both by optical imaging and magnetic resonance velocimetry in various geometries, is also supported by numerical results with no-slip boundary conditions \cite{Varnik:2003,Xu:2005}. However recent studies have suggested more complicated scenarii where the nature of the yielding transition may depend on the interactions between the individual components \cite{Becu:2006} or on the confinement of the system \cite{Isa:2007,Goyon:2008} pointing to the absence of any universal local relationship between the shear stress and the shear rate in the vicinity of yielding.

{\it Boundary conditions} may be another crucial control parameter for the yielding transition. Indeed, it has long been known that rheological measurements can strongly depend on wall surface roughness due to the presence of apparent wall slip \cite{Barnes:1999}. Whether a soft glassy material may or may not slip at the walls is thus expected to affect the solid--fluid transition under shear. So far local measurements investigating the influence of wall roughness in the context of yielding have only shown a weak impact of slip on the nature of the transition both in microfluidic devices \cite{Isa:2007,Goyon:2008} and in macroscopic gaps \cite{Meeker:2004}.

In this Letter the yielding transition of a Laponite suspension is probed within a 1~mm gap for two different boundary conditions (rough and smooth walls) using standard rheology, time-resolved local velocity measurements, and direct optical imaging simultaneously. As previously observed, rough walls lead to shear localization without significant temporal fluctuations. However smooth walls provide an entirely different scenario where the initially solidlike material breaks up into a very heterogeneous pattern of macroscopic solid domains separated by fluidized zones. These solid pieces progressively erode leading to complete fluidization of the sample. A tentative interpretation of such a striking impact of boundary conditions on yielding is suggested in light of recent works on cooperativity in soft glassy materials.

Laponite powder (Rockwood, grade RD), a synthetic clay made of platelets of diameter 25--30~nm and thickness 1--2~nm, is dispersed at 3~wt.~$\%$ in ultrapure water \cite{Bonn:2002,DiLeonardo:2005a}. Hollow glass spheres of mean diameter 6~$\mu$m (Sphericel, Potters) are added at 0.3~wt.~$\%$ to the Laponite suspension and act as contrast agents for ultrasonic velocimetry (see below). Within 30~min of magnetic stirring, the dispersion becomes homogeneous and very viscous but remains fluid. Due to the presence of the large glass particles, the solution is slightly turbid. The elastic modulus $G'$ slowly builds up and, after approximately two hours, overcomes the loss modulus $G"$, corresponding to ``solidification.'' In our samples, $G'$  and $G"$ obtained from oscillatory measurements in the linear regime show no significant frequency dependence in the range 0.01--10~Hz. Over a two week period, $G'$ and $G"$ vary from 200 to 1000~Pa and from 15 to 40~Pa respectively due to aging \cite{Bonn:2002,DiLeonardo:2005a}. Meanwhile, the yield stress measured from stress sweep experiments at 1~Hz shows a variation from 30 to 100~Pa.

In the following, our purpose is to focus on the influence of boundary conditions on yielding. We compare the flow behavior of two Laponite samples in a sand-blasted and in a ``smooth'' Plexiglas Couette cell (rotating inner cylinder radius $R=24$~mm, gap width 1~mm, and height 30~mm). The respective roughnesses of these cells measured from atomic force microscopy are 0.6~$\mu$m and 15~nm, which will be referred to as ``rough'' and ``smooth'' \cite{Roughness}. A rheometer (TA Instruments AR1000N) imposes a constant shear rate ($\gammap\simeq 20$~s$^{-1}$) and measures the corresponding shear stress $\sigma$ as a function of time $t$. Velocity profiles accross the gap are recorded every 0.5~s at about 15 mm from the cell bottom using ultrasonic speckle velocimetry, a technique based on acoustic tracking of scatterers (here, the hollow glass spheres) \cite{Manneville:2004}. Direct visualization of the sheared sample is performed using a CCD camera (Cohu 4192). The temperature is kept constant at $25\pm 0.1$\degC. Before any measurement, the Laponite sample is pre-sheared for 1~min at +1500~s$^{-1}$ and for 1~min at -1500~s$^{-1}$ to erase most of the sample history. We then proceed with a standard oscillatory test at 1~Hz in the linear regime for 2~min. We checked that this procedure leads to reproducible results over a few hours.

\begin{figure}[htb]
\begin{center}
\scalebox{1}{\includegraphics{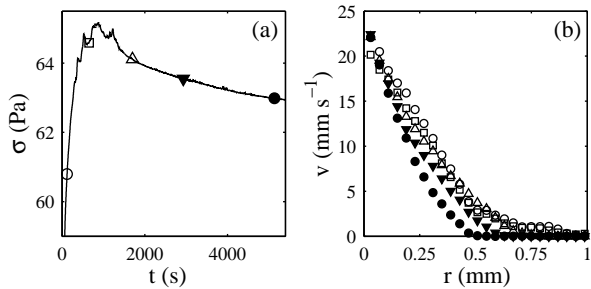}}
\end{center}
\vspace{-0.5cm}
\caption{Flow behavior in a rough Couette cell after a constant shear rate $\gammap=25$~s$^{-1}$  is imposed at $t=0$~s. (a) Stress response $\sigma$ vs time $t$. (b) Velocity profiles $v(r)$ at $t=123$~s ($\circ$), $653$~s ($\square$), $1702$~s ($\vartriangle$), $2931$~s ($\blacktriangledown$), and $5137$~s ($\bullet$). $r$ is the radial distance from the inner rotating cylinder. Error bars are of the order of the marker size.}
\label{fig1}
\end{figure}

Figure~\ref{fig1} shows the results obtained in the {\it rough} Couette cell after a step-like shear rate $\gammap=25$~s$^{-1}$ is imposed at $t=0$. The velocity profiles present shear localization: an unsheared solidlike region close to the stator coexists with a sheared fluid region on the rotor side. The sample remains optically homogeneous during the whole experiment. The size of the unsheared band slowly increases with time for $t\gtrsim 1500$~s, consistently with the slow decrease of $\sigma(t)$, so that the system has not completely reached a stationary state after more than 5000~s. For a fixed shearing duration, we checked that the size of the unsheared band decreases with the imposed shear rate
(data not shown \cite{Gibaud}). All these observations are consistent with previous results obtained in the absence of wall slip \cite{Coussot:2002,Moller:2008,Rogers:2008} and will serve here as a hallmark of the ``standard'' yielding transition.

\begin{figure}[htb]
\begin{center}
\scalebox{1}{\includegraphics{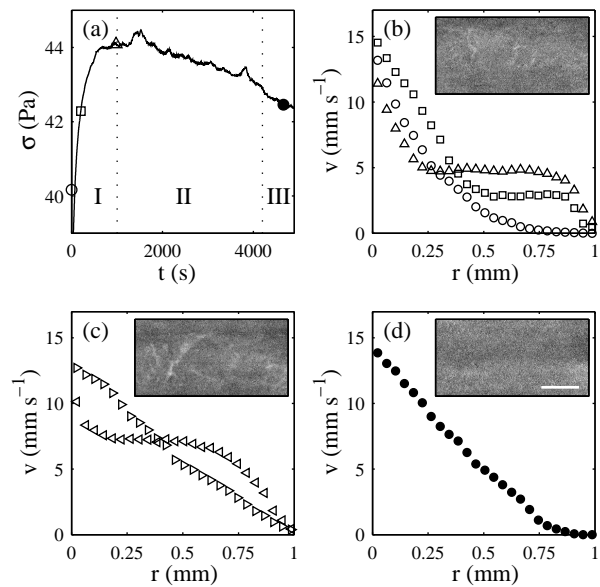}}
\end{center}
\vspace{-0.5cm}
\caption{Flow behavior in a smooth Couette cell after a constant shear rate $\gammap=17$~s$^{-1}$  is imposed at $t=0$~s. (a) Stress response $\sigma$ vs time $t$. Velocity profiles $v(r)$: (b) in regime I at $t=3$~s ($\circ$), $205$~s ($\square$), and $980$~s ($\vartriangle$); (c) in regime II at $t=2235$~s ($\vartriangleleft$) and $2245$~s ($\vartriangleright$); (d) in regime III at $t=4665$~s ($\bullet$). Insets in (b), (c), and (d): pictures of the sample in regime I, II, and III at $t=874$~s, $2236$~s, and $4350$~s respectively. The white bar corresponds to 5~mm. See also~\cite{SupMat}.}
\label{fig2}
\end{figure}

\begin{figure}[htb]
\begin{center}
\scalebox{1}{\includegraphics{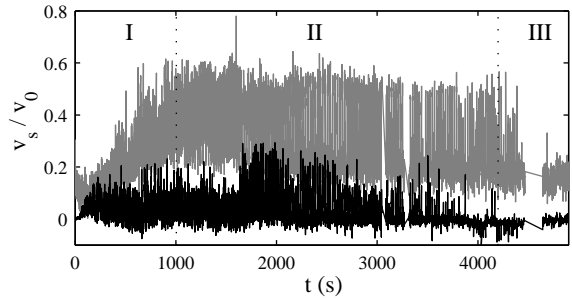}}
\end{center}
\vspace{-0.5cm}
\caption{Normalized apparent slip velocities $v_s/v_0$ in the smooth geometry derived from linear fits of the velocity profiles over 0.2~mm from the stator (black) and from the rotor (grey), where $v_0$ is the rotor velocity.}
\label{fig3}
\end{figure}

Measurements performed in the {\it smooth} Couette cell for a similar start-up experiment are gathered in Fig.~\ref{fig2}. The stress response of Fig.~\ref{fig2}(a) is qualitatively very similar to that of Fig.~\ref{fig1}(a). One can only note that noisy features are detectable in $\sigma(t)$ until at least $t\simeq 4000$~s whereas, in the rough geometry, $\sigma(t)$ slowly and smoothly relaxes after a much shorter noisy transient. Surprisingly the velocity profiles in the smooth geometry (Fig.~\ref{fig2}(b)--(d)) reveal a totally different yielding behavior with much more complex spatiotemporal dynamics where three regimes may be distinguished. For the first few seconds ($\circ$ in Fig.~\ref{fig2}(b)), the velocity profiles present an unsheared region close to the stator, similar to that observed in the rough geometry, and the sample appears homogeneous. However, in the smooth cell, the solidlike region soon detaches from the stator ($\square$ and $\vartriangle$ in Fig.~\ref{fig2}(b)): the velocity of the unsheared region progressively increases as the width of the sheared layer close to the rotor decreases. As shown in Fig.~\ref{fig3}, this corresponds to a strong increase of apparent wall slip at the rotor. This slow process, noted regime I in Fig.~\ref{fig2}(a), ends at $t\simeq 1000$~s when the unsheared region has invaded almost the whole gap and undergoes a plug-like flow at a velocity $v_{\rm solid}$ which roughly corresponds to half the rotor velocity. Direct observations of the sample during regime I indicate the slow apparition of heterogeneities in the sample, most probably due to local variations of the hollow glass spheres concentration. At the end of regime I, these heterogeneities are arranged in a fracture-like pattern \cite{SupMat}.

Regime II, that extends from $t\simeq 1000$~s to $t\simeq 4200$~s, is characterized by a flow pattern that is strongly heterogeneous in both space and time. Indeed, as illustrated in Fig.~\ref{fig2}(c), velocity measurements display alternately plug-like flow in most of the gap and viscous flow profiles ($\vartriangleleft$ and $\vartriangleright$ respectively). Consequently slip velocities show huge fluctuations both at the rotor and at the stator (see Fig.~\ref{fig3}). Note that one velocity profile corresponds to an average over $\Delta t=0.22$~s whereas the rotation period of the rotor is 8.8~s. Therefore we conclude that the sample is constituted of macroscopic solid parts that are separated by fluid regions and rotate at a velocity $v_{\rm solid}\simeq 7$\mms. This picture is confirmed by direct optical imaging of the sample \cite{SupMat}, which reveals large domains of typical size 1~mm--1~cm carried away by the flow and whose edges appear brighter than the background in the inset of Fig.~\ref{fig2}(c).

\begin{figure}[htb]
\begin{center}
\scalebox{1}{\includegraphics{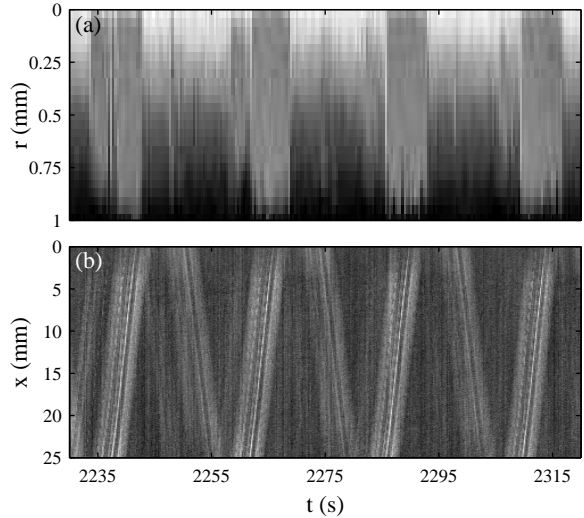}}
\end{center}
\vspace{-0.5cm}
\caption{Comparison of local velocimetry and optical imaging in regime II of the experiment shown in Fig.~\ref{fig2}.
 (a) Spatiotemporal diagram of the velocity $v(r,t)$ coded in linear grey levels from 0 (white) to 17\mms\ (black). Plug-like profiles appear as uniform grey patches. 
(b) Spatiotemporal diagram of the intensity $I(x,t)$ along a horizontal line of the pictures of the sample (see insets in Fig.~\ref{fig2}) taken at the height of the ultrasonic transducer (about 15~mm from the cell bottom). A linear grey scale is used. The edges of solid pieces appear brighter than the background. Since the CCD camera and the ultrasonic transducer used for velocimetry are at 90\degre, the time axis for $I(x,t)$ was shifted by $T/4\simeq 5.7$~s.}
\label{fig4}
\end{figure}

Figure~\ref{fig4} provides a closer look into regime II by comparing simultaneous velocimetry and imaging. The spatiotemporal diagram of Fig.~\ref{fig4}(a) shows a clear alternance of plug-like and viscous flow profiles with a period $T\simeq 23$~s. The same period is observed in Fig.~\ref{fig4}(b) from optical imaging and corresponds to the rotation period $T=2\pi R/v_{\rm solid}$ of solid pieces carried away by the flow. $v_{\rm solid}$ can also be independently recovered from the slope of the traces left by the solid domains in Fig.~\ref{fig4}(b) \cite{Note}. Finally, it is directly checked that viscous flow coincides with homogeneous parts of the images whereas plug-like profiles are associated to heterogeneities in the detected intensity. Therefore the complex temporal behavior of the velocity measurements in regime II is merely a consequence of the highly heterogeneous distribution of the solid pieces in the azimuthal direction.

\begin{figure}[htb]
\begin{center}
\scalebox{1}{\includegraphics{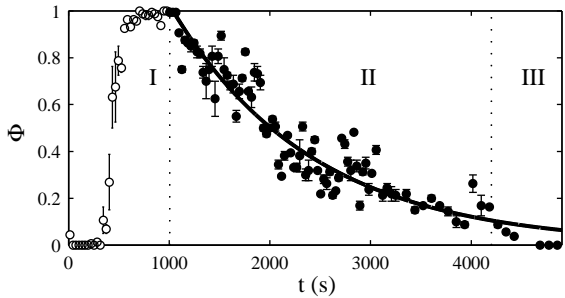}}
\end{center}
\vspace{-0.5cm}
\caption{Fraction $\Phi$ of plug-like velocity profiles measured within a constant time window of 50~s during the experiment shown in Fig.~\ref{fig2}. A velocity profile is counted as ``plug-like'' when the local shear rate in the middle of the gap is smaller than the threshold value 8~s$^{-1}$. The thick line is the best exponential fit of $\Phi(t)$ over regime II which yields a characteristic relaxation time of 1450~s.}
\label{fig5}
\end{figure}

On longer time scales, it can be seen from the images that the fraction of solid pieces in the sample slowly decreases thoughout regime II \cite{SupMat}. More quantitatively, Fig.~\ref{fig5} shows the temporal evolution of the fraction $\Phi$ of plug-like profiles during the whole experiment: $\Phi=0$ ($\Phi=1)$ corresponds to a homogeneous fluid (solid) state. The slow relaxation of $\Phi(t)$ in regime II suggests that the solid domains get eroded by the surrounding sheared fluid. At the end of regime II, no solid part is left and the sample becomes optically homogenous again (see inset of Fig.~\ref{fig2}(d)). Velocity profiles in regime III are stationary and linear except for a small arrested band of width 120~$\mu$m close to the stator (see Fig.~\ref{fig2}(d)). It is important to notice that this asymptotic flow behavior is {\it not} compatible with that obtained in the rough geometry (where half the gap is in a fluid state) even if one accounts for the difference in apparent shear rates. Finally, in the smooth geometry, we found that the scenario of fragmentation and fluidization described above occurs for imposed shear rates ranging from 10 to 65~s$^{-1}$. At lower shear rates, this scenario becomes hardly accessible as the time scale for fragmentation diverges and at larger shear rates, the sample becomes fluid within seconds and remains so. More quantitative results on the characteristic relaxation time for $\Phi(t)$ and its dependence on the imposed shear rate will be provided elsewhere \cite{Gibaud}.

In summary we have shown that the yielding properties of a soft glassy material may depend drastically on boundary conditions. In the absence of wall slip, shear localization is observed, whose very slow dynamics can most probably be attributed to aging \cite{Rogers:2008}. When wall slip is allowed, our system, namely a Laponite suspension seeded with hollow glass spheres, shows a much richer behavior characterized by the breaking up of the initially solid material followed by a progressive fluidization through slow erosion of the solid fragments. This scenario leads us to revisit some recent observations where the lower temporal and/or spatial resolutions of the velocimetry techniques did not allow for definite conclusions. More precisely, puzzling oscillations of velocity measurements were reported in Laponite samples at low shear rates \cite{Moller:2008,Ianni:2008}. Another example is provided by some glassy colloidal star polymers where intermittent jammed states were recorded together with strong wall slip \cite{Holmes:2004}. We believe that such dynamics are the signature of the fragmentation process unveiled in the present study which may thus be very general as soon as apparent slip is involved. A deeper investigation of the interplay between the material microstructure and the wall surface properties, e.g. by varying the Laponite and/or the glass spheres concentration, will be performed to provide more insight on the generality of this behavior.

Our main conclusion is that boundary conditions may have a drastic influence on the nature of the yielding transition. In the case of slippery walls, a possible interpretation is that breaking up in regime I starts due to some temporary sticking of the sample to the cell walls. This is supported by the fact that the fluctuations of slip velocities increase throughout regime I. During such sticking events, the solid may accumulate stresses that eventually overcome locally the yield stress and give birth to a fluidized zone which may subsequently erode the surrounding solid due to viscous stresses (regime II). This interpretation raises a few important issues.

(i) What parameters control the competition between fragmentation and erosion? For instance, we observed that the age of the 
sample has a strong impact on the relative durations of regimes I and II: samples older than the ones investigated here present a larger density of solid pieces at the end of regime I and a faster fluidization \cite{SupMat}.

(ii) In the case where fragmentation dominates, the typical size of solid pieces constitutes a {\it macroscopic} characteristic length $\lambda$ which is difficult to assess quantitatively with the present setup. Is $\lambda$ an intrinsic parameter of the material? Very recently, Goyon {\em et al.} \cite{Goyon:2008} evidenced the existence of a ``cooperativity length'' $\xi$ in confined flows of concentrated emulsions that characterizes the influence region of rearrangements. $\xi$ is non-zero only in the jammed material and falls in the 10~$\mu$m range ($\sim$~5 droplet diameters). How may $\lambda$ be related to $\xi$?

(iii) What is the relevant combination between material properties and surface roughness that triggers stick rather than slip? Does the roughness characteristic length provide a cutoff for $\lambda$ \cite{Steinberger:2007,Bocquet:2007}? All these questions not only urge for more experimental effort but also for theoretical and numerical approaches that could account for the interaction between a solid boundary and a jammed structure close to yielding.

\begin{acknowledgments}
The authors wish to thank P.~Jop, L.~Petit, and N.~Taberlet for their help on the system and on the experiment, as well as A.~Piednoir and M.~Monchanin for the surface roughness measurements.
\end{acknowledgments}



\begin{thebibliography}{40}

\bibitem{Liu:1998}
A. Liu and S. R. Nagel,
Nature {\bf 396}, 21 (1998).

\bibitem{Trappe:2001}
V. Trappe {\em et al.},
Nature {\bf 411}, 772 (2001).


\bibitem{Barnes:1999}
H. A. Barnes,
J. Non-Newtonian Fluid Mech. {\bf 56}, 221 (1995); {\bf 70}, 1 (1997); {\bf 81}, 133 (1999).

\bibitem{Pignon:1996}
F. Pignon {\em et al.},
J. Rheol. {\bf 40}, 573 (1996).

\bibitem{Coussot:2002}
P. Coussot {\em et al.},
Phys. Rev. Lett. {\bf 88}, 218301 (2002).

\bibitem{Rogers:2008}
S. A. Rogers {\em et al.},
Phys. Rev. Lett. {\bf 100}, 128304 (2008).

\bibitem{Moller:2008}
P. C. F. M{\o}ller {\em et al.},
Phys. Rev. E {\bf 77}, 041507 (2008).

\bibitem{Varnik:2003}
F. Varnik {\em et al.},
Phys. Rev. Lett. {\bf 90}, 095702 (2003).


\bibitem{Xu:2005}
N. Xu {\em et al.},
Phys. Rev. Lett. {\bf 94}, 016001 (2005).

\bibitem{Becu:2006}
L. B{\'e}cu {\em et al.},
Phys. Rev. Lett. {\bf 96}, 138302 (2006).

\bibitem{Isa:2007}
L. Isa {\em et al.},
Phys. Rev. Lett. {\bf 98}, 198305 (2007).

\bibitem{Goyon:2008}
J. Goyon {\em et al.},
Nature {\bf 454}, 84 (2008).

\bibitem{Meeker:2004}
S. P. Meeker {\em et al.},
Phys. Rev. Lett. {\bf 92}, 198302 (2004).

\bibitem{Bonn:2002}
D. Bonn {\em et al.},
Phys. Rev. Lett. {\bf 89}, 015701 (2002).

\bibitem{DiLeonardo:2005a}
R. Di Leonardo {\em et al.},
Phys. Rev. E {\bf 71}, 011505 (2005).

\bibitem{Roughness}
These estimates correspond to the average standard deviation of height profiles taken from various AFM images.

\bibitem{Manneville:2004}
S.~Manneville {\em et al.}, Eur. Phys. J. AP {\bf 28}, 361 (2004).

\bibitem{Gibaud}
T. Gibaud {\em et al.},
in preparation.

\bibitem{SupMat}
See EPAPS Document No. [to be inserted] for movies of sheared samples of different ages.

\bibitem{Note}
Since the sample is semi-transparent and the rotor is made of Plexiglas, solid pieces are observed on both sides of the Couette cell when they pass in front of and behind the rotor. Therefore, stripes with respectively upward and downward slopes show up in the spatiotemporal diagram of Fig.~\ref{fig4}(b).

\bibitem{Ianni:2008}
F. Ianni {\em et al.},
Phys. Rev. E {\bf 77}, 031406 (2008).

\bibitem{Holmes:2004}
W. H. Holmes {\em et al.},
J. Rheol. {\bf 48}, 1085 (2004).

\bibitem{Steinberger:2007}
A. Steinberger {\em et al.},
Nature Materials {\bf 6}, 665 (2007)

\bibitem{Bocquet:2007}
L. Bocquet and J.-L. Barrat,
Soft Matter {\bf 3}, 685 (2007)

\end{thebibliography}
\end{document}